\documentclass[aps,pre,twocolumn,superscriptaddress,showpacs,floatfix]{revtex4-1}

\usepackage{amsfonts,amssymb,latexsym,xspace,epsfig,graphics,color}
\usepackage{amsmath,enumerate,stmaryrd,xy,stackrel}

\usepackage{tikz}
\usepackage{tkz-berge}
\usetikzlibrary{arrows,snakes,backgrounds}
\usetikzlibrary{shapes}
\definecolor{pinegreen}{cmyk}{0.92,0,0.59,0.25}
\definecolor{royalblue}{cmyk}{1,0.50,0,0}
\definecolor{lavander}{cmyk}{0,0.48,0,0}
\definecolor{violet}{cmyk}{0.79,0.88,0,0}
\tikzstyle{cblue}=[circle, draw, thin,fill=black, scale=0.5]
\tikzstyle{qgre}=[circle, draw, thin,fill=black, scale=0.5]
\tikzstyle{rpath}=[thick, black, opacity=0.4]
                        \usetikzlibrary{positioning}
                          
\usepackage{subfigure}

\begin{document}


\title{Comment on ``Nodal infection in Markovian susceptible-infected-susceptible and susceptible-infected-removed epidemics on networks are non-negatively correlated''}


\author{Pablo M. Rodriguez}
\email[]{pablor@icmc.usp.br}
\thanks{Corresponding author}
\affiliation{Instituto de Ci\^encias Matem\'aticas e de Computa\c{c}\~ao, Universidade de S\~ao Paulo.\\ Caixa Postal 668, 13560-970 S\~ao Carlos, SP, Brazil.}

\author{Alejandro Rold\'an-Correa}
\email[]{alejandro.roldan@udea.edu.co}
\thanks{}
\affiliation{Instituto de Matem\'aticas, Universidad de Antioquia.\\ Calle 67 Nro. 53-108, Medell\'in, Colombia.}

\author{Leon Alexander Valencia}
\email[]{lalexander.valencia@udea.edu.co}
\thanks{}
\affiliation{Instituto de Matem\'aticas, Universidad de Antioquia.\\ Calle 67 Nro. 53-108, Medell\'in, Colombia.}


\date{\today}

\begin{abstract}
Cator and Van Mieghem [Cator E, Van Mieghem P., Phys. Rev. E 89, 052802 (2014)] stated that the correlation of infection at the same time between any pair of nodes in a network is non-negative for the Markovian SIS and SIR epidemic models. The arguments used to obtain this result rely strongly on the graphical construction of the stochastic process, as well as the FKG inequality. In this note we show that although the approach used by the authors applies to the SIS model, it cannot be used for the SIR model as stated in their work. In particular, we observe that monotonicity in the process is crucial for invoking the FKG inequality. Moreover, we provide an example of simple graph for which the nodal infection in the SIR Markovian model is negatively correlated.

\end{abstract}


\maketitle


\section{Introduction}

The understanding of the correlation structure in epidemic-like processes on networks is an issue of important research. The interest in this subject rely on its direct applicability in the modeling of phenomena like the propagation of an infectious disease or a piece of information on a population. Two classical models are usually taken as a basis for the formulation of mathematical and computational epidemic-like processes; namely, the SIS (susceptible-infected-susceptible) and the SIR (susceptible-infected-removed) models. 

The purpose of this note is to comment about the results communicated by Cator and Van Mieghem in \cite{cator}, where it is stated that the correlation of infection at the same time between any pair of nodes in a network is non-negative for the Markovian SIS and SIR epidemic models. The arguments used in \cite{cator} rely on the graphical construction of the SIS stochastic process, as well as an elegant application of the FKG inequality to a discrete-time version of the original process. In the same work, the authors claim that the developed techniques can be adapted for the SIR stochastic process. However, we show that the conclusion for the SIR model cannot be obtained from the arguments developed in \cite{cator}. Indeed, we will notice that a key for these techniques is the existence of monotonicity in the process according to a partial order introduced by the authors, something that it does not hold for the SIR model. Moreover, we provide a counter-example given a simple graph for which the nodal infection is negatively correlated for the original Markovian process, while it is positively correlated if one consider an encoding coming from a function of the original process.

The note is organized as follows. In Section 2 we review the description of the SIR stochastic process through its graphical construction, we briefly present the main idea behind the discrete-time construction given by \cite{cator}, and we discuss the fail into try to apply the FKG inequality for this process. Section 3 is devoted to our counter-example.

\section{The SIR stochastic process}

The SIR stochastic model on a finite network $G=([N],E)$ may be formulated as a continuous-time Markov chain $(X_t)_{t\geq 0}$ with states space given by $\mathcal{S}:=\{0,1,2\}^{[N]}$. For any time $t\geq 0$, a state of the process is a function $X_t : [N]\rightarrow \{0,1,2\}$, where for any node $i$ the random variable $X_t(i)$ denotes the infectious state of node $i$ at time $t$. Here $X_t(i)=0$ means not infected, $X_t(i)=1$ means infected, and $X_t(i)=2$ means removed. If the process is in state $X$ at a given time $t\geq 0$ then, the state of a node $i\in [N]$ evolves according to the following transitions and rates:
\begin{equation}\label{eq:trans1}
\mathbb{P}(X_{t+h}(i)=1|X_t(i)=0)= \lambda \,h\, n_1(i,X)+ o(h),
\end{equation}
\begin{equation}\label{eq:trans2}
\mathbb{P}(X_{t+h}(i)=2|X_t(i)=1)= \delta \,h\, + o(h),
\end{equation}
where $n_1(i,X)$ denotes the number of neighbors of nodes $i$ at state $1$, $\lambda >0$ denotes the infection rate from infected nodes to susceptible nodes, $\delta >0$ denotes the recovery rate of infected nodes, and $o(h)/h\rightarrow 0$ as $h\rightarrow 0$. We point out that the choice of $0,1$ and $2$ to represent the different states of the process is necessary in order to well define the stochastic SIR model as a Markovian process. In addition, we opt for this representation instead of the usual $S, I$ and $R$ representation (see for example \cite{van}) with the purpose of discussing the applicability of the construction given by \cite{cator} to the SIR Markovian process.

It is a well-known fact that spatial correlations difficult an exact analysis of the SIR model on a network. Indeed, the usual approach to deal with this kind of processes is the mean-field approximation, where it is assumed that there are no dynamical correlations at first order. In other words, the mean-field approach assumes that the expected values of variable pairs factorize, i.e. $\mathbb{E}\left(X_t(i)X_t(j)\right)=\mathbb{E}\left(X_t(i)\right)\mathbb{E}\left(X_t(j)\right)$, for $i,j\in [N]$ and $t>0$. If the considered process is a $\{0,1\}$-valued process, the previous equality implies $\mathbb{P}\left(X_t(i)=1,X_t(j)=1\right)=\mathbb{P}\left(X_t(i)=1\right)\mathbb{P}\left(X_t(j)=1\right)$, which is quite interesting for applications. In \cite{cator} the authors claim that for any finite network the inequality 
\begin{equation}\label{eq:result}
\mathbb{E}\left(X_t(i)X_t(j)\right)\geq 
\mathbb{E}\left(X_t(i)\right)\mathbb{E}\left(X_t(j)\right)
\end{equation} holds for the SIS and for the SIR Markovian processes, which is the same to say that the random variables are non-negatively correlated. In the sequel we show why this fact cannot be concluded from their arguments.

\subsection{The Harris' graphical construction}

\begin{figure}[!h]
\begin{center}

\includegraphics{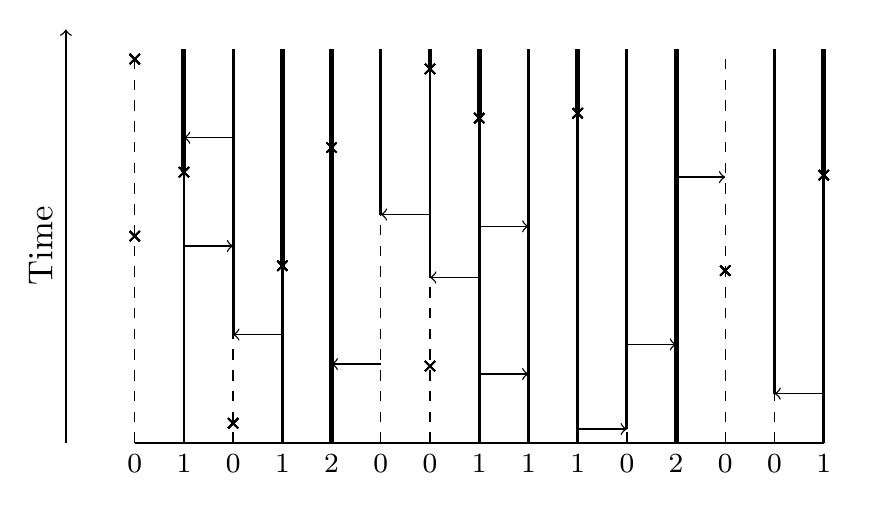}

\caption{Realization of the SIR model on a path graph. The susceptible, infected and removed states are drawn in dashed, thin and thick lines, respectively. The thin arrows indicate the times at which a susceptible node may becomes infected, and the crosses the times at which an infected node may becomes removed.}\label{FIG:harris}
\end{center}
\end{figure}

We consider the graphical representation of the SIR model, which is well-known with the name of Harris' graphical construction \cite{Harris}.  Consider the independent collections of independent point Poisson processes $\{\mathcal{N}_{(i,j)}^{\lambda}: i,j\in [N] \text{ and } i\sim j\}$ and $\{\mathcal{N}_{i}^{\delta}: i\in [N]\}$ with intensities $\lambda$ and $\delta$, respectively. At each arrival time of $\mathcal{N}_{(i,j)}^{\lambda}$ if nodes $i$ and $j$ are in states $1$ and $0$, respectively, then the state of node $j$ is updated to $1$. On the other hand, at each arrival time of $\mathcal{N}_{i}^{\delta}$, if node  $i$ is in state $1$, then $i$ changes to state $2$. In this way we obtain a version of the SIR model on $G$ with transitions and rates given by \eqref{eq:trans1} and \eqref{eq:trans2}. See Fig. \ref{FIG:harris} for an illustration of a possible realization of the model through this construction. In the sequel we call any point coming from the process $\mathcal{N}_{(i,j)}^{\lambda}$ an infection arrow, and any point coming from the process $\mathcal{N}_{i}^{\delta}$ a cure mark. For further details on the Harris'graphical representation of interacting particle systems see for instance Durrett \cite{durrett}.\\

The graphical construction allows, for example, to construct realizations of different stochastic processes through the same families of point Poisson processes. As we will see next, it allows us to show that there is non-monotonicity for the SIR stochastic process in the construction proposed by \cite{cator}.

\subsection{The Cator-Van Mieghem construction and the non-monotonicity for the SIR model}

Inspired by the Harris' graphical construction, Cator and Van Mieghem~\cite{cator} give a description of the entire evolution of the SIS process through all times in $[0,t]$, for $t>0$. The only difference with the SIR model is that if there is a cure mark at time $s$ in node $i$, and $i$ is at state $1$ then its states changes to $0$. For the sake of  completeness we briefly write here the construction of the process introduced in \cite{cator}, which we call the Cator-Van Mieghem construction. We refer the reader to that paper for further details. The first step is to define a random function $Z:W\to\mathcal{M},$ where $W$ is the collection of all nodes and all directed edges in the graph and $\mathcal{M}$ is the collection of all finite subsets of $[0,t].$ The function $Z$ is defined by mean of the involved point processes in such a way that the following holds: if $s\in Z(e)$  for an edge $e=(i,j)\in E,$ then at time $s$ the node $i$ will attempt to infect the node $j.$ On the other hand, if $s\in Z(i)$ for a node $i\in [N],$ then the node $i$ will heal at time $s$ (see \cite[Section II.A]{cator}). Once $Z$ is suitably defined the next step consist into construct a discrete time version of the SIS process. For this, choose $n$ equally spaced time points in $[0,t]$; i.e., consider $t_1,\ldots,t_n$ with $t_k:=kt/n$, and define the vector $X^{(n)}(t_k)\in\{0,1\}^{[N]}$ describing the state of the (discrete-time) SIS process at time $t_k.$ In order to accomplish this, it is necessary to consider some auxiliary functions and random variables. Take first the function  $Z_n:W\times\{t_1,\ldots,t_n\}\to \{0,1\}$ in such a way that
    $$Z_n(i,t_k):=\left\{
    \begin{array}{ll}
    1 & \text{if } Z(i)\cap (t_{k-1},t_k]=\emptyset,\\
    0 & \text{if } Z(i)\cap (t_{k-1},t_k]\neq\emptyset,
    \end{array}
    \right.$$
for any $i\in [N]$, and 
   $$Z_n(e,t_k):=\left\{
  \begin{array}{ll}
  0 & \text{if } Z(e)\cap (t_{k-1},t_k]=\emptyset,\\
  1 & \text{if } Z(e)\cap (t_{k-1},t_k]\neq\emptyset,
  \end{array}
  \right.$$
  for any $e\in E$.  It is not difficult to see that the space $\mathcal{X}_n:=\{0,1\}^{W\times\{t_1,\ldots,t_n\}}$ has a natural partial ordering: 
let $Z_1, Z_2\in \mathcal{X}_n,$ then $Z_1\leq Z_2$ whether 
$$Z_1(i,t_k)\leq Z_2(i,t_k)$$ 
and
$$Z_1(e,t_k)\leq Z_2(e,t_k).$$

Now define the intermediate state random vector $Y^{(n)}(t_k)\in\{0,1\}^{[N]}$, where
$$ Y_i^{(n)}(t_k):=X_i^{(n)}(t_{k-1})Z_n(i,k)$$ and finally take $X_i^{(n)}(t_k)$ equals to
$$Y_i^{(n)}+[1-Y_i^{(n)}(t_k)]\max_{j:(i,j)\in E}Y_j^{(n)}(t_k)Z_n((i,j),t_k).$$
In order to make the dependence on $Z$ explicit, $X^{(n)}(t_k)$ is also denoted by $X_i^{(n),Z}(t_k).$  The essential part of this construction is that for any $t\geq 0$
\begin{equation}\label{eq:essFKG}
X_i^{(n),Z_1}(t)\leq X_i^{(n),Z_2}(t)
\end{equation} provided $X_i^{(n),Z_1}(0)=X_i^{(n),Z_2}(0)$ and $Z_1\leq Z_2.$ That is, $X_i^{(n),Z}(t)$, regarded as a function of $Z$, is an increasing function on the partially ordered set $\mathcal{X}_n$. This is the property which allows to use the FKG inequality and therefore to conclude (\ref{eq:result}) by a limit argument, see \cite[Section III]{cator}

According to Cator and Van Mieghem~\cite{cator} the same construction could be used to prove  (\ref{eq:result}) for the SIR model. Moreover, the authors claim that the same function $Z$ does the work, with the only difference that if an infected node heals at time $s$, its state is must be changed to the recovered state. Unfortunately, the construction is only valid when the process is monotone which is not the case for the SIR process as we see in a simple example in Fig. 2. Indeed, in Fig. 2 we have two realizations, namely $Z_1$ and $Z_2$, such that $Z_1\leq Z_2$ and $X^{(n),Z_1}(0)=X^{(n),Z_2}(0)=(0,1,0,0)$ but $X^{(n),Z_1}(t)=(0,2,0,0)$ and $X^{(n),Z_2}(t)=(0,1,1,0)$. Then, even if starting from the same initial configuration and assuming $Z_1\leq Z_2$, the respective states of the processes at time $t$ form two non-comparable configurations. This means that the Cator-Van Mieghem construction it does not work for the SIR model, and therefore it is not possible to use the FKG inequality with this approach.  

Actually, although the Cator-Van Mieghem construction gives a novel proof of (\ref{eq:result}) for the SIS epidemic process, and there is not doubt that it is an interesting and useful construction to study monotone processes, we point out that \eqref{eq:result} it has already been proved for any monotone process, see~\cite{Harris2,donnelly}.

At this point one can ask about the validity of \eqref{eq:result} for the SIR model. In the next Section we provide a counter-example of this claim when one consider the $X_{i}(t)$'s random variables of the original Markovian SIR process and, at the same time, we discuss about what happens whether one consider an encoding of the states obtained as a function of the original process.   

\begin{figure}[!h]\label{example}
\begin{center}
\subfigure[][]{

\includegraphics{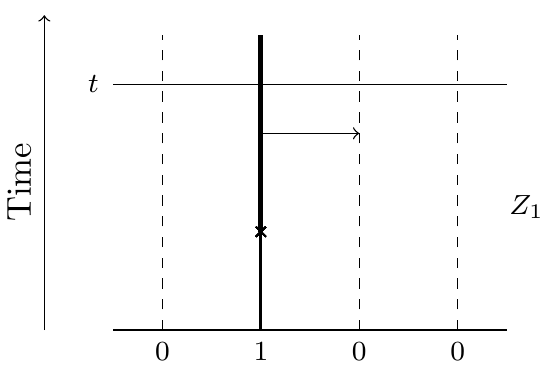}

}

\subfigure[][]{

\includegraphics{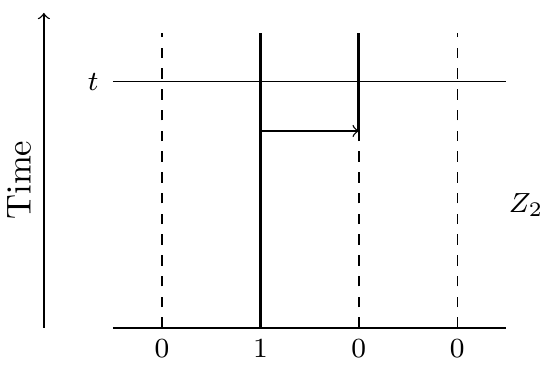}

}

\end{center}

\caption{Comparison between two realizations of the SIR process according to two functions $Z_1$ and $Z_2$ such that $Z_1 \leq Z_2$. (a) Realization of the SIR process according to the $Z_1$ function. (b) Realization of the SIR process according to the $Z_2$ function.}\label{FIG:example}
\end{figure}



\section{A counter-example}

Let's take the path graph with two nodes; i.e., $G=([2],E)$, where $E=\{(1,2)\}$, and consider the SIR model on $G$ with infection rates given by $\lambda=2$ and cure rates given by $\delta =1$. For the sake of simplicity lets calculate the covariance between the states of nodes $1$ and $2$ at $t=1$ provided $X_0(1)=1$ and $X_{0}(2)=0$. That is, consider $X_1(1)$ and $X_1(2)$. The joint distribution of these random variables can be obtained by observing the behavior of the marks of the Poisson processes $\mathcal{N}_{(1,2)}^{\lambda}$, $\mathcal{N}_{1}^{\delta}$, and $\mathcal{N}_{2}^{\delta}$ in the interval $[0,1]$. For instance, conditioned on the event  $\{X_0(1)=1,X_{0}(2)=0\},$
$$\mathbb{P}\left(X_1(1)=1,X_1(2)=0\right)=e^{-1}\,e^{-2},$$
because in order to maintain the states of both nodes at time $1$ the process cannot have marks of cure in $1$, and no infection arrows from $1$ to $2$, during the time window $[0,1]$. Note that this is the same to say that the event $\{S_1^{\delta}>1,S_{(1,2)}^{\lambda}>1\}$ occurs, where $S_i^{\delta}$ is the instant of occurrence of the first point of $\mathcal{N}_{i}^{\delta}$, $i=1,2$, while $S_{(1,2)}^{\delta}$ is the instant of occurrence of the first mark of $\mathcal{N}_{(1,2)}^{\lambda}$. With a little more work we can obtain, again conditioned on  $\{X_0(1)=1,X_{0}(2)=0\},$ that $\mathbb{P}\left(X_1(1)=1,X_1(2)=1\right)$ is equal to
$$\int_{0}^{\infty}\mathbb{P}\left(X_1(1)=1,X_1(2)=1 | S_{(1,2)}^{\lambda} =t \right)\, f_{S_{(1,2)}^{\lambda}}(t)\,dt,$$
and since $\mathbb{P}\left(X_1(1)=1,X_1(2)=1 | S_{(1,2)}^{\lambda} =t \right)$ is equals to $
e^{-1}\,e^{-(1-t)}\, {\bf{1}_{\{0<t<1\}}}$, and $f_{S_{(1,2)}^{\lambda}}(t)=2 \,e^{-2t}\, {\bf{1}_{\{t>0\}}}$ is the density probability function of an exponencial random variable, we conclude that
$$\mathbb{P}\left(X_1(1)=1,X_1(2)=1\right)= 2\,e^{-2}\,(1-e^{-1}).$$

Analogously, we calculate
\begin{eqnarray*}
\mathbb{P}\left(X_1(1)=1,X_1(2)=2\right)&=& e^{-3}\,(e-1)^2,\\
\mathbb{P}\left(X_1(1)=2,X_1(2)=0\right)&=& \frac{1-e^{-3}}{3},\\
\mathbb{P}\left(X_1(1)=2,X_1(2)=1\right)&=& e^{-3}\,(e-1)^2,\\
\mathbb{P}\left(X_1(1)=2,X_1(2)=2\right)&=& \frac{2\,e^{-3}\,(e-1)^{-3}}{3}. 
\end{eqnarray*}

Thus, by a straight calculation we obtain 
$$\text{COV}(X_1(1),X_1(2))=-\frac{1}{3 e^4} - \frac{2}{e^3} + \frac{5}{e^2} - \frac{8}{3 e}<0,$$
or, in other words,
$$\mathbb{E}\left(X_1(1)\, X_1(2)\right)< 
\mathbb{E}\left(X_1(1)\right)\mathbb{E}\left(X_1(2)\right).
$$

We have showed that by considering the Markovian SIR process $(X_t)_{t\geq 0}$ neither the Cator-Van Mieghem construction applies, nor the inequality \eqref{eq:result} may be guarantee. On the other hand, we notice that by considering the random variables given by $Y_t(i):=1$ whether $X_{t}(i)=1$ and $Y_t(i):=0$, whether $X_{t}(i)\neq 1$, for any $i\in [N]$ and $t\geq 0$, one may show that $\text{COV}(Y_1(1),Y_1(2))>0$. This suggest that although \eqref{eq:result} may be true for the process $(Y_t)_{t\geq 0}$, this is not longer a Markovian process. In order to see that, by returning to the counter-example, observe that for some times $0\leq t_0<t_1<t_2$ we have, on one hand
$$\mathbb{P}(Y_{t_2}(2)=1\, | \, Y_{t_1}(2)=0,\, Y_{t_0}(2)=1)=0,$$
while, on the other hand
$$\mathbb{P}(Y_{t_2}(2)=1\, | \, Y_{s}(2)=0,\, \text{for all } 0\leq s\leq t_1)>0.$$
In words, we have that the past of the stochastic process influences its future so the Markovian property it is not satisfied. The null probability above may be obtained by noticing that, conditioned on the event $\{Y_{t_1}(2)=0,\, Y_{t_0}(2)=1\}$, the event $\{X_{t_1}(2)=2\}$ occurs and then  $\{Y_{t_2}(2)=1\}$ it is not longer possible for $t_2 >t_1$. On the other hand, the occurrence of the event $\{Y_{s}(2)=0,\, \text{for all } 0\leq s\leq t_1\}$ may implies the occurrence of the event $\{X_{s}(2)=0,\, \text{for all } 0\leq s\leq t_1\}$ which guarantee that the event $\{Y_{t_2}(2)=1\}$ has a positive probability to occur, provided $t_2 >t_1$.

We conclude that further research with new arguments must be developed to study correlations for nodal infection in general graphs. Hence, understanding the correlation structure for SIR epidemic models remain as an interesting open problem for further consideration.

\smallskip
\begin{acknowledgments}
This work was supported by CNPq (304676/2016-0), FAPESP (2016/11648-0, 2017/10555-0) and Universidad de Antioquia. Part of this work was carried out during a stay of the first author at the Universidad de Antioquia; he is grateful for their hospitality and support. The authors are grateful to the reviewers for their comments and suggestions.
\end{acknowledgments}


\begin{thebibliography}{99}

\bibitem{cator}
E. Cator, and P. Van Mieghem, Phys. Rev. E 89, 052802 (2014).

\bibitem{donnelly}
P. Donnelly, Math Biosci. 117(1-2), 49-75 (1993).

\bibitem{durrett}
R. Durrett, Ten Lectures on Particle Systems. {Lecture Notes in Mathematics} {1608}. Springer, New York (1995).

\bibitem{Harris2}
{T. Harris,} 
{Ann. Probab.} {5} (3), 451-454 (1977).

%
\bibitem{Harris}
{T. Harris,}
{Adv. Math.} {9}, 66--89 (1972).
%


\bibitem{van}
P. Van Mieghem, F. D. Sahnehz, and C. Scoglioz, 53rd IEEE Conference on Decision and Control, Los Angeles, CA, 2014, 6228-6233.


\end{thebibliography}

\end{document}